# Accretion Flows around Black Holes

Ramesh Narayan
Harvard-Smithsonian Center for Astrophysics
60 Garden Street, Cambridge MA 02138

August 7, 1995

## 1 Introduction

Relativistic stars such as black holes and neutron stars play a central role in modern astrophysics, witness the number of articles on the subject in these proceedings. The study of accretion flows around relativistic stars is an especially exciting branch of astrophysics, as it includes both black holes and neutron stars and spans objects ranging from stellar mass accretors to supermassive black holes.

In this article we concentrate on accreting black holes. A small but rapidly growing subset of X-ray binaries (XRBs) has been identified as black hole candidates. A few of these candidates like Cyg X-1, A0620-00, V404 Cyg, Nova Muscae 1991 (Cowley 1994, van Paradijs & McClintock 1995) have good mass determinations which show that the compact accreting stars in these systems are too massive to be neutron stars and therefore must be black holes. The remaining black hole candidates are identified purely on spectral similarity to the well-established candidates (e.g. White 1994). A second class of candidate black holes are active galactic nuclei (AGNs), which are believed to be accreting supermassive black holes (mass $\sim 10^5 - 10^9 M_\odot$, see the article by Martin Rees in this volume for a detailed discussion of AGNs). In a few cases like NGC4258 (Miyoshi et al. 1995), and Sgr A* at the center of our Galaxy (Genzel & Townes 1987, Genzel, Hollenbach & Townes 1994), the evidence for a black hole is based on dynamical information (see Scott Tremaine, this volume, for other examples). However, for the majority of AGNs, no such direct evidence exists, though extremely plausible theoretical arguments can be made (e.g. Rees 1984).

## 2 X-rays and $\gamma$-rays from Accreting Black Holes

The problem we focus on in this article is the fact that accreting black holes almost always emit a substantial fraction of their luminosity in hard X-rays and $\gamma$-rays, roughly in the energy range 10 keV to a few ×100 keV. Their spectra in this energy range can be described as a power-law with a photon index $\alpha_N$:

$$N_E dE \propto E^{-\alpha_N} dE, \qquad (1)$$

where $N_E dE$ represents the number of photons per second in the photon energy range $E$ to $E + dE$. The quantity $E^2 N_E \propto E^{(2-\alpha_N)}$ is a useful way of characterizing the spectrum as it represents the energy emitted by the source per logarithmic interval of E. Generally, spectra with $\alpha_N < 3$ may be considered to be hard spectra; spectra with $\alpha_N < 2$ are exceptionally hard since they have most of the energy coming out in the hardest photons.

The observational evidence suggests that all black hole XRBs spend at least part of their time in a hard state (the so-called Low State, see §4.2) with $\alpha_N \sim 1.5 - 2.5$ (Grebenev et al. 1993, Harmon et al. 1994, Gilfanov et al. 1995). The power-law rolls over above $E \sim$ few $\times$ 100 keV in most cases, but there may be an excess above the continuum at MeV energies in a few sources (e.g. Cyg X-1). AGNs too appear to have a more or less universal hard component in their spectra. The data are unfortunately less complete at this time (since bright AGNs, although $\sim 10^8$ times more luminous than bright XRBs, are $\sim 10^5$ times farther away and therefore $\sim 10^2$ times fainter.) The spectra of a number of quasars, Seyferts and other AGNs have been measured up to $E \sim 10-20$ keV using the Einstein Observatory, Ginga and other X-ray satellites (Williams et al. 1992, Elvis et al. 1994a). At these energies the spectra are power-laws with $\alpha_N \lesssim 2$. The quasar S5 0014+813, at redshift 3.384, has been observed with ASCA to have an $\alpha_N = 1.6$ power-law extending up to at least 40 keV in the rest frame of the source (Elvis et al. 1994b). The OSSE instrument on the Compton Gamma-Ray Observatory (CGRO) has been studying several AGNs at energies between 50 keV and 10 MeV and has found power-laws with $\alpha_N \sim 2$ extending above 100 keV in many cases (Maisack et al. 1993, Johnson et al. 1994, Kinzer et al. 1994).

These observations place strong constraints on theoretical models. The radiating gas has to be optically thin in order to produce a power-law spectrum, and the electron temperature has to be $\gtrsim 10^9$ K in order to produce photons up to 100 keV. The requirements are quite difficult to meet with accretion models.

The most common model of accretion flows is the thin accretion disk model developed by Shakura & Sunyaev (1973), Novikov & Thorne (1973) and Lynden-Bell & Pringle (1974). In this model, the accreting gas is assumed to be cool, with gas temperature $T_{\text{gas}}$ much less than the virial temperature $T_{\text{vir}}$ ($\sim$ few $\times 10^{12} K/r$, where $r$ is the radius in Schwarzschild units). Since the gas is cool, (i) radial pressure forces are negligible and the angular velocity $\Omega$ of the gas is essentially equal to the Keplerian value, $\Omega_K = (GM/R^3)^{1/2}$, and (ii) the gas forms a thin disk with its vertical thickness $H$ being much less than the radius. The energy equation, which describes the balance between the local energy generation due to viscous dissipation and the cooling due to vertical radiative transfer and radiation from the surface of the disk, simplifies considerably in the thin disk geometry and it is straightforward to calculate the density and temperature of the accreting gas as a function of $r$. The thin disk model has been applied successfully to accreting white dwarfs and pre-main-sequence stars (Frank, King & Raine 1992).

The thin accretion disk model suffers from one major uncertainty which plagues all studies of accretion, namely the unknown nature of viscosity in the gas. Microscopic viscosity is much too small to produce significant accretion. Various macroscopic mechanisms such as shear-driven hydrodynamic turbulence, convection and MHD instabilities have been explored, the last process being considered particularly attractive (Balbus & Hawley 1991). In models of accretion flows it is usual to absorb the uncertainty in the viscosity mechanism into a dimensionless parameter $\alpha$ (Shakura & Sunyaev 1973) and to write the kinematic coefficient of viscosity $\nu$ as

$$\nu = \alpha c_s H \sim \frac{\alpha c_s^2}{\Omega_K}, \tag{2}$$

where $c_s$ is the sound speed of the gas. In certain circumstances, it is necessary to modify equation (2) to enforce causality (Popham & Narayan 1992, Narayan 1992, Syer & Narayan 1993, Narayan, Loeb & Kumar, 1994, Kato & Inagaki 1994). We do not discuss these questions further, although they do represent major unsolved problems in accretion physics.

The problem we focus on here is the fact that the thin accretion disk model, despite being plausible and self-consistent, is unable to explain the hard spectra of accretion flows around black holes. Thin accretion disks around black holes are nearly always optically thick in the vertical direction and therefore produce blackbody-like spectra (eg. Frank et al. 1992). This is of course in conflict with the power-law spectra observed. Further, the effective temperature is much too low:

$$T_{\text{eff}} \sim (6 \times 10^7 K) m^{-1/4} \dot{m}^{1/4} r^{-3/4}, \tag{3}$$

where $m$ is the mass of the black hole in solar units and $\dot{m}$ is the mass accretion rate in Eddington units ($\dot{M}_{\text{Edd}} = 2.2 \times 10^{-8} m \ M_\odot \text{yr}^{-1}$, assuming an efficiency of 0.1, cf Frank et al. 1992). Equation (3) shows that with reasonable parameters it is impossible to obtain a temperature anywhere close to $T \sim 10^9$ K. Moreover, the thin disk model predicts that $T_{\text{eff}}$ should decrease with increasing black hole mass, whereas observations indicate that the characteristic temperature of the radiation is $\gtrsim 10^9$ K for both stellar-mass and supermassive black holes.

Before proceeding to discuss other accretion models, two comments are in order. First, even though the thin disk model cannot explain power-law X-ray/$\gamma$-ray emission, it does explain softer components in the spectra of accreting black holes. Black hole XRBs in their "High" and "Very High" states (cf. §4.2) have soft spectra with $kT \sim 0.7 - 1$ keV. This emission could very well be produced by a standard thin disk (e.g. Ebisawa 1994). Similarly, quasars have a "blue bump" in their spectra which is very likely due to thin disk emission (Malkan & Sargent 1982).

The second comment is that all the discussion in this article pertains to emission between 10 keV and 1 MeV. In addition to this emission, which is often thermal, a completely different kind of emission has been seen in blazars (a particularly active variety of AGNs). A number of these sources have been observed by the Egret detector on CGRO to produce tremendous amounts of radiation at GeV − TeV energies (Hartman et al. 1994). This emission appears to be relativistically beamed and seems to be produced by nonthermal mechanisms associated with relativistic jets. This topic is not reviewed here.

# 3 Hot Accretion Flow Models

In view of the inability of the thin accretion disk model to produce hard X-rays and $\gamma$-rays, various attempts have been made to develop alternate models of accretion. A common feature of these models is that the radiating electrons are relativistic, so that Compton-scattering is important. In a seminal paper, Sunyaev & Titarchuk (1980, see also Titarchuk 1994) discussed Comptonization of soft photons by hot electrons and showed that, for optical depths $\sim$ a few, power-law spectra extending up to $E \sim kT$ are obtained. The soft seed photons for the Comptonization could come from a variety of sources — photons from a standard thin accretion disk, photons from the surface of the central star (in the case of accretion onto a neutron star), or synchrotron photons produced locally in the hot magnetized plasma. At the temperatures of interest, electron-positron pair processes can be important, and there is an extensive literature on calculating these effects (Svensson 1982, 1984, Zdziarski 1985, White & Lightman 1989). Most discussions are in terms of thermal models, but nonthermal models have also been considered at times (Svensson 1994). Recent observations have confirmed earlier indications that some black hole XRBs and most Seyferts have turnovers in their spectra between $E \sim 100$ keV and 1 MeV. This turnover poses severe difficulties for purely nonthermal models (e.g. Zdziarski et al. 1993, Fabian 1994). Therefore, the focus presently is on thermal models, with perhaps a fraction of the emission coming from nonthermal processes.

Three main categories of hot accretion flow models have been considered in the literature.

## 3.1 Corona Models

In this model, the hot gas is postulated to be in a corona above a standard thin accretion disk. The corona is characterized by its density (or equivalently optical depth) and temperature (or Compton $y$-parameter). The disk provides the soft photons for Comptonization. X-ray data on AGNs and black hole XRBs show spectral features such as iron emission lines and a bump between 10 and 30 keV which indicates X-ray reflection in cool gas (Pounds et al. 1990, Matsuoka et al. 1990). These features are explained in the corona model as due to reflection of the hot coronal radiation off the underlying cool disk (Guilbert & Rees 1988, Lightman & White 1988, Fabian 1994 and references therein).

Haardt & Maraschi (1991, 1993) have developed an improved corona model where the temperature is not a free parameter but is solved for by balancing the viscous heating rate of the gas and the Compton cooling rate by soft photons from the disk. The disk emission is calculated self-consistently by including irradiation from the corona. The model assumes that most of the viscous energy dissipation occurs in the corona (as argued by Field & Rogers 1993) rather than in the disk, though most of the mass accretion is assumed to occur in the cool disk.

Coronal models though promising are still incomplete at this time since the coronal gas has little or no dynamics. The density and temperature of the corona are often adjusted freely, and even in the approach of Haardt & Maraschi the density is left unspecified. The models need to be developed further so that the density of the coronal gas is calculated self-consistently as a function of radius.

## 3.2 SLE Two-Temperature Model

In an important paper, Shapiro, Lightman & Eardley (1976, SLE) showed that, in addition to the standard thin disk model, a second solution to the accretion flow equations is permitted at somewhat sub-Eddington accretion rates. The gas in this solution is optically thin and quite hot. SLE proposed the important idea that the accreting plasma may be two-temperature close to the accreting star, with the ions being significantly hotter than the electrons. The electrons, which are cooled primarily by Comptonization of soft photons, achieve temperatures $\sim 10^8 - 10^9$ K.

The optically thin nature of the gas in the SLE solution, plus the high temperature, are precisely the features one needs to explain the high energy spectra of accreting black holes. Therefore, the SLE solution has been widely applied in models of XRBs and AGNs (e.g. Kusunose & Takahara 1985, 1989, White & Lightman 1989, Wandel & Liang 1991, Luo & Liang 1994, Melia & Misra 1993).

A major problem with the SLE solution is that the gas is thermally unstable (Piran 1978, Wandel & Liang 1991, Narayan & Yi 1995b). If the temperature of the electrons in an equilibrium flow is slightly perturbed, the gas either cools catastrophically to the thin disk solution or heats up catastrophically without bound. The instability operates on the thermal timescale ($\sim 1/\alpha\Omega_K$), which is much faster than the accretion timescale.

## 3.3 Optically-Thin Advection-Dominated Model

Recently, a new class of optically thin hot solutions has been discovered (Narayan & Yi 1994, 1995ab, Abramowicz et al. 1995, Chen 1995, Chen et al. 1995) which have all the advantages of the SLE solution without suffering from a violent thermal instability. A crucial feature of these solutions is that most of the viscously dissipated energy is advected with the accreting gas as stored entropy, with only a small fraction of the energy being radiated. The flows are therefore described as being *advection-dominated*. The accreting gas in these solutions is optically thin and is in principle thermally unstable. However, because of advection-domination the radiative cooling is only a small perturbation on the overall energetics of the gas, and further the accretion flow timescale is quite short. For both reasons, the thermal instability does not have time to grow to any significant amplitude and the flow is effectively stable to thermal perturbations (cf. Kato, Abramowicz & Chen 1995). The flows are, however, convectively unstable (Narayan & Yi 1994, 1995a, Igumenshchev, Chen & Abramowicz 1995).

The idea of advection-dominated accretion goes back to Begelman (1978) and Begelman & Meier (1982) who suggested that at very high $\dot{m}$ radiation may be trapped in the accreting gas and be unable to escape. Solutions corresponding to this regime were obtained by Abramowicz et al. (1988). These flows are optically thick and not very hot, and therefore are not of interest for the discussion in this article.

Optically thin advection-dominated flows operate in a different regime where the gas density is so low that the radiative efficiency is very poor. In this case, energy is advected not because the radiation is trapped, but rather because the heated gas radiates very little of its internal energy.

The net effect is the same, however, namely that most of the binding energy is retained in the gas as thermal energy and is carried into the black hole. Early work on these flows was done by Rees, Begelman, Blandford & Phinney (1982, who called the gas configuration an "ion torus," see also Phinney 1981), Matsumoto, Kato & Fukue (1985), and Narayan & Popham (1993, who found optically-thin advection-dominated flows in accretion disk boundary layers at low $\dot m$). However, it is only recently that the properties of these hot flows have been studied in detail (Narayan & Yi 1994, 1995ab, Abramowicz et al. 1995, Chen 1995, Chen et al. 1995). Assuming the plasma becomes two-temperature, Narayan & Yi (1995b) found that the electron temperature lies in the range $10^9 - 10^{10}$ K, depending on $\dot m$. The ion temperature on the other hand is nearly virial ($\sim 10^{12} \text{K}/r$) because of the low radiative efficiency. The accreting gas is therefore almost spherical in shape rather than disk-like or toroidal (Narayan & Yi 1995a). Despite this, the gas does rotate (with sub-Keplerian $\Omega$) and has angular momentum transport and viscous dissipation just as in regular thin disks. These features, plus the subsonic radial velocity (except close to the horizon), distinguish this model from pure spherical accretion models (e.g. Ipser & Price 1983, Melia 1994).

The optically-thin advection-dominated model requires the gas to be fairly optically thin, which means that the solutions exist only for relatively low accretion rates, $\dot m \lesssim 0.3\alpha^2$. In addition, because of the strong advection, the radiative efficiency is low. For both reasons, these solutions work best when applied to low luminosity sources. For such sources, the model has been quite successful.

Figure 1 shows an advection-dominated model (Narayan, Yi & Mahadevan 1995) of Sagittarius A* (Sgr A*), the putative supermassive black hole at the center of our Galaxy (Genzel & Townes 1987, Genzel et al. 1994). This model, which corresponds to a black hole mass of $7 \times 10^5 M_\odot$ and an accretion rate of $\dot M \sim 10^{-5}\alpha M_\odot \text{yr}^{-1}$, nicely reproduces the observed spectrum of the source over nearly ten decades of frequency. Furthermore, the model is able to explain the extremely low X-ray luminosity of the source, $\sim 10^{37} \text{ergs s}^{-1}$, despite making use of a fairly substantial $\dot M$. This is welcome because there is evidence for substantial amounts of gas in the vicinity of Sgr A* (Genzel & Townes 1987, Melia 1992), and until now it has been a mystery (Genzel et al. 1994) why the source is as dim as it is. An unusually low luminosity is of course natural with an advection-dominated flow since most of the energy disappears into the black hole.

Figure 2 shows another application, this time to the black hole soft X-ray transient A0620-00 in its quiescent state. In this case, the model consists of a standard thin accretion disk on the outside, extending down to a radius $r = 4400$, and an optically thin advection-dominated flow on the inside extending down to the horizon (Narayan, McClintock & Yi 1995). The model is able to explain the spectrum of the source quite well, fitting both the X-ray and the optical/UV observations.

The hot advection-dominated solutions discussed here are currently the only known thermally stable accretion flows which achieve the kind of relativistic temperatures needed to explain the X-ray and $\gamma$-ray spectra of accreting black holes. For this reason, there is considerable effort under way to explore the full potential of these models. A currently open question is whether this model, which by its low $\dot m$ and low radiative efficiency is best suited for low-luminosity sources, can be applied also to objects with near-Eddington luminosities. Figure 3 shows the limit of validity of the hot advection-dominated solutions in the $r\dot{m}$ plane. For reasonable values of $\alpha$, say $\lesssim 0.3$, the hot solutions are present only for $\dot{m} \lesssim 0.03$ (the lower solid line in the upper panel of Fig. 3). Only if $\alpha \sim 1$ (Fig. 3, lower panel) can these solutions be pushed to interesting values of $\dot{m}$ appropriate to luminous sources (Narayan 1995, Narayan, Yi & Mahadevan 1996). An unsolved issue is whether $\alpha$ can be this large. Another problem is that only solutions which are very close to the upper limits shown in Fig. 3 have high efficiencies. Therefore these models will need some measure of fine-tuning if they are to be applied to bright sources. It is always possible to modify the model of viscous dissipation and radiation in such a way as to achieve higher radiative efficiency even at low values of $\dot{m}$, but it remains to be seen if such a modification can be done in a physically plausible way. Other unresolved issues center on whether two-temperature plasmas are allowed in hot low density flows (see Begelman & Chiueh 1988, and Appendix A of Narayan & Yi 1995b), on what effect electron-positron pair production will have on the nature of the solutions, and on the role of the inner boundary condition near the horizon.

## 4 Directions for Future Research

The ultimate goal in the study of black hole accretion is easily stated:

*Given the mass $M$ of an accreting black hole, the mass accretion rate $\dot{M}$, and a few additional parameters (say the disk inclination $i$, black hole specific angular momentum $a$), we would like to be able to calculate all the details of the flow and be able to predict the spectrum and variability of the source.*

The thin accretion disk model does satisfy this goal in the sense that it gives a clear prediction for the spectrum of a source as a function of system parameters. Further, there has been interesting work on thermal instabilities which seems to explain X-ray nova outbursts in soft X-ray transients (Huang & Wheeler 1989, Mineshige & Wheeler 1989, but see Lasota 1995). Unfortunately, as we have seen, the thin disk model does not produce any hard radiation.

The most promising direction to pursue at this time is to try and develop the hot advection-dominated solutions further and to look for variants of these solutions which may be better suited to explain high luminosity sources. Orienting such theoretical work toward explaining the rich phenomenology that X-ray and $\gamma$-ray satellites are currently providing could be quite rewarding.

### 4.1 Unresolved Theoretical Issues

Perhaps what is needed most urgently is a systematic search for all stable accretion flow equilibria. Figure 3, based on Chen et al. (1995), shows the inter-relationship among the three stable equilibria which are currently known, viz. the standard thin disk model (Shakura & Sunyaev 1973), the optically-thick advection-dominated model (Abramowicz et al. 1988), and the optically-thin advection-dominated model (Narayan & Yi 1994, 1995b, Abramowicz et al. 1995, Chen 1995). An important question is whether there could be other equilibria if we relax some of the assumptions made by Chen et al. (1995). For instance, in addition to the two-temperature models described by Narayan & Yi (1995b), are there single-temperature models where the ions and electrons are strongly coupled by non-thermal processes? How would these models differ from two-temperature models? If we allow the electrons to have some degree of nonthermal energy distribution, how much will the results change? Can there be important nonlocal effects, such as Comptonization of radiation emitted at one radius by gas at another radius, and can this open up new global equilibria which go beyond the local analyses done so far? These are important questions to which current research is being directed.

Once a fairly complete set of equilibria has been identified, two important issues will need to be resolved:

(1) It is possible to have multiple equilibria for given values of $\dot m$ and $r$; as Fig. 3 shows, this happens already just with the three known equilibria. Consider a system at relatively low $\dot m$ in region 3 of Fig. 3. Such a system has to choose between the standard thin disk solution and the hot advection-dominated solution. Which equilibrium does it choose? There are clearly several possibilities. Some systems may exist purely in the cool disk state, while others (for example Sgr A*, Fig. 1) may exist entirely in the hot state. Another possibility is that a system may exist in the cool state at some radii and in the hot state at other radii; our model of A0620-00 in quiescence (Fig. 2) is such an example. Yet another possibility is that both solutions may be simultaneously present at the same radius; for example, part of the accretion may happen along an equatorial thin disk, and part may happen in a hot advection-dominated corona. Such a model would be very close in spirit to the corona models described earlier except that when fully developed this model would presumably include a full calculation of the vertical structure of the gas and would allow us to calculate how much mass goes into the corona and how the gas dynamics out there works. A final possibility, of course, is that the system may go into a time-dependent mode where it switches from one state to another on some characteristic time scale.

(2) For some ranges of $\dot m$ and $r$ there may be no stable equilibrium at all. Region 4 in Fig. 3 corresponds to this situation. What will an accreting system that falls within region 4 do? It seems obvious that the flow will be forced into a time-dependent mode (Chen et al. 1995), perhaps oscillating between different equilibria in such a way as to satisfy the required $\dot m$ in the mean. The possibilities here are very rich and the time-dependent dynamics could be very interesting. Observationally, we note that most bright black hole accretors show strong variability both in their luminosity and spectrum.

## 4.2  Clues from Observations of Black Hole XRBs

Although the number of AGNs known is vastly larger than the number of black hole XRBs, the latter are likely to prove more useful for understanding black hole accretion. Because XRBs are much closer to us than AGNs they tend to be brighter and to provide better data. Also, because the time scales in XRBs are significantly shorter than those in AGNs, it is possible to obtain considerable information on variability and fluctuations of XRBs, whereas equivalent data on AGNs would require observations over prohibitively long periods of time.

Black hole XRBs exhibit four distinct states, distinguished by luminosity and spectral shape (e.g. Tanaka 1989, Grebenev et al. 1993, van der Klis 1994, Gilfanov et al. 1995). In order of increasing luminosity, these states are:

(1) *Off/Quiescent State*: This is seen predominantly in X-ray novae between outbursts. In at least two sources, V404 Cyg and A0620-00, low-level X-ray emission has been seen in quiescence, and presumably all quiescent sources have such emission. The X-ray luminosity is in the range $L_X \sim 10^{30} - 10^{34}$ ergs s$^{-1} \ll L_{\rm Edd}$, where $L_{\rm Edd} = 10^{38} m$ ergs s$^{-1}$ is the Eddington luminosity. The nature of the spectrum is not known, but according to the advection-dominated model (Fig. 2) the spectrum is fairly hard. Sgr A$^*$ (Fig. 1) may be considered an example of a galactic nucleus in the Off/Quiescent State.

(2) *Low State*: In this state, the spectrum is exceptionally hard, typically $\alpha_N \sim 1.5 - 2$. The luminosity is typically $L_X \sim 10^{36} - 10^{37.5}$ ergs s$^{-1} \lesssim 0.01 L_{\rm Edd}$.

(3) *High State*: The luminosity is greater than in the Low State, $L_X \sim 0.01 - 0.9 L_{\rm EDD}$, and the spectrum is very soft. Indeed, at most about 10% of the emission is above 10 keV.

(4) *Very High State*: This state, first discovered in the source GX 339-4 by Ginga observations (Miyamoto et al. 1991), corresponds to the highest luminosity, $L_X \gtrsim 0.9 L_{\rm Edd}$. The spectrum consists of an ultrasoft component which contains most of the luminosity plus a hard tail with $\alpha_N \sim 2.5$.

The existence of so many distinct states is doubtless a very important clue for understanding the nature of accretion flows around black holes. If we accept the advection-dominated model for the Quiescent State, an obvious question is: could the same model be applied also to the Low State? The hard spectrum of the Low State implies a very hot optically thin flow, and perhaps the only difference between this state and the Off State is the value of $\dot{m}$. The High State appears to be a good candidate for the standard thin accretion disk model since the observed temperature is consistent with that predicted by equation (3), with perhaps some modifications due to electron scattering (Ebisawa 1994). Why is there an abrupt change between the Low and High States? Perhaps the boundary between the two states corresponds to the limiting $\dot{m}$ of the hot advection-dominated solution. This proposal works quite well if $\alpha \sim 1$ (Narayan 1995, Narayan et al. 1996), and so the question is whether such a large value of $\alpha$ is physically reasonable. What is the nature of the Very High State? If the $\dot{m}$ in the High State is already too large to allow a hot advection-dominated flow, then the $\dot{m}$ in the Very High State certainly is much too large. Yet this state is seen to have a hard tail in the spectrum. What could produce the tail? One might speculate that the hard radiation in this case is produced by a corona. Since the typical value of $\alpha_N$ in the Low State is smaller than that in the Very High State (van der Klis 1994) it is reasonable to propose different origins for the two hard components. Finally, does the upper optically-thick advection-dominated solution branch (see Fig. 3) play a role in the Very High State?

An interesting correlation discussed by Gilfanov et al. (1995) is the fact that the spectral index $\alpha_N$ in the Low State varies considerably from one system to another but is relatively independent of the hard X-ray luminosity in any single system. What determines the value of $\alpha_N$? The fact that $\alpha_N$ is independent of luminosity suggests that $\dot{M}$ may not be the key, though one should keep in mind that the hard X-ray luminosity is not necessarily directly proportional to $\dot{M}$. Nor is $\alpha_N$ likely to depend on the black hole mass, since XRBs and AGNs have similar spectra in their hard states even though they differ by orders of magnitude in mass. Therefore, there has to be a third parameter, which may be the inclination $i$ of the system or the specific angular momentum $a$ of the black hole.

Black hole XRBs tend to be variable on a wide range of time scales and some systems even display quasi-periodic oscillations. In general, it appears that the largest variability is seen in the hard spectral components, while the soft components remain relatively steady. This could well be an important clue to the nature of the hot gas in these systems. The reader is referred to van der Klis (1994), Kouveliotou (1994) and Gilfanov et al. (1995) for more details on the variability of black hole XRBs.

## 4.3 Black Holes versus Neutron Stars

An interesting sub-topic in high energy accretion is the distinction between flows around black holes and those around weakly magnetized neutron stars (e.g. van der Klis 1994). (X-ray pulsars are excluded from this discussion since the neutron star magnetic field plays an important role in these systems and the spectral properties are very different.) There are two reasons why this subject is of interest. First, if the differences between neutron stars and black holes are understood well, then one could hope to develop better techniques to distinguish black hole XRBs from neutron star XRBs in the observations. This will make it easier to identify new black hole systems. Second, if the differences between the two kinds of systems can be fitted naturally into a theoretical framework, then we can have confidence in our general understanding of high energy accretion flows.

Black holes and neutron stars both have deep relativistic potentials, which is of course why the two systems look so similar when they accrete. The main distinction between the two cases is that the inner boundary conditions are different.

1. *Radiation Boundary Condition*: In the case of a black hole, the accreting matter disappears through the horizon, carrying its energy with it. In contrast, a neutron star has a surface which re-radiates whatever energy flows in. Sunyaev et al. (1991ab, see also Liang 1993) have argued that Compton-cooling of the accretion flow by stellar photons will make the spectra of neutron star XRBs steeper than those of black hole XRBs. Narayan & Yi (1995b) confirmed this with detailed calculations of hot two-temperature advection-dominated flows around neutron stars. The observational situation is that hard tails do occur in neutron star systems in their low intensity states (Barret & Grindlay 1995, Barret et al. 1996). Fitting these tails with power laws generally yields spectral indices $\alpha_N \gtrsim 3$, though on one occasion the X-ray burster in Terzan 2 showed a very hard spectrum with $\alpha_N = 1.7$ (Barret et al. 1991). Thermal bremsstrahlung fits give temperatures $kT \sim 30 - 40$ keV (Barret & Vedrenne 1994). Therefore, on average, neutron star hard tails are cooler as well as steeper/softer than black hole tails (which have $\alpha_N \sim 1.5 - 2.5$, $kT \gtrsim 100$ keV). As further evidence that neutron star flows are cooler, we note that electron-positron annihilation features have been seen only in black hole candidates (1E1740.7-2942, Bouchet et al. 1991; Nova Muscae 1991, Goldwurm et al. 1992).

2. *Flow boundary condition*: The second difference, which has been hardly explored, follows again from the fact that the black hole has a horizon whereas the neutron star has a hard surface. The accreting gas responds to these two boundaries differently. In the case of the black hole, the gas makes a sonic transition and falls into the horizon supersonically. In contrast, gas flowing onto a neutron star has to slow down, either subsonically (e.g. Popham & Narayan 1992) or following a shock (Kluzniak & Wagoner 1985), and settle on the surface. It is conceivable that this difference leads to important changes in the gas dynamics close to the center. One might speculate for instance that the nature of outflows and jets depends very sensitively on the inner boundary condition. Many AGNs have relativistic jets (Zensus & Pearson 1987, Begelman, Blandford & Rees 1984) and it is now becoming clear that many XRBs too have such ejections. The best examples of XRB jets are all found in black hole candidates, e.g. 1E1740.7-2942 (Mirabel et al. 1992), GRS 1915+105 (Mirabel & Rodriguez 1994a), GRO J1655-40 (Tingay et al. 1994, Hjellming & Rupen 1995), GRS 1758-258 (Mirabel & Rodriguez 1994b). The only XRBs which have jets and which might be neutron stars are SS 433 (Margon 1984) and Cyg X-3 (Molnar, Reid & Grindlay 1988). Both of these are nonstandard XRBs and may well be accreting black holes. If this is the case, then one could say that relativistic jets are an unambiguous signature of black holes. Alternatively, since SS 433 and Cyg X-3 have somewhat slower moving jets (relativistic $\beta \sim 0.2 - 0.3$) compared to the other cases, one could say that the speed of the jet is determined by the boundary condition in such a way that settling flows make slower jets (M. J. Rees, private communication). One point to keep in mind is that neutron star X-ray bursters probably have not been imaged in the radio to the same sensitivity level as the black hole candidates. Such observations need to be done before one can draw any firm conclusions on the differences between outflows from neutron stars and black holes.

# 5 Conclusion

This article has tried to show that black hole accretion is an open field with many opportunities and many unsolved problems. The field currently lacks an accepted paradigm, though the recent work on optically-thin advection-dominated accretion flows seems promising. The present time may be particularly good for an all-out assault in this area because of the wealth of observational clues pouring in from high energy space missions like CGRO and (very soon) XTE. With the INTEGRAL mission set to be launched at the turn of the century we can anticipate much more data in the years ahead.

The author is grateful to the following colleagues and collaborators for comments and instruction:

Didier Barret, Josh Grindlay, Jean-Pierre Lasota, Rohan Mahadevan, Jeff McClintock, Martin Rees, and Insu Yi. This work was supported in part by grant NAG 52837 from NASA.

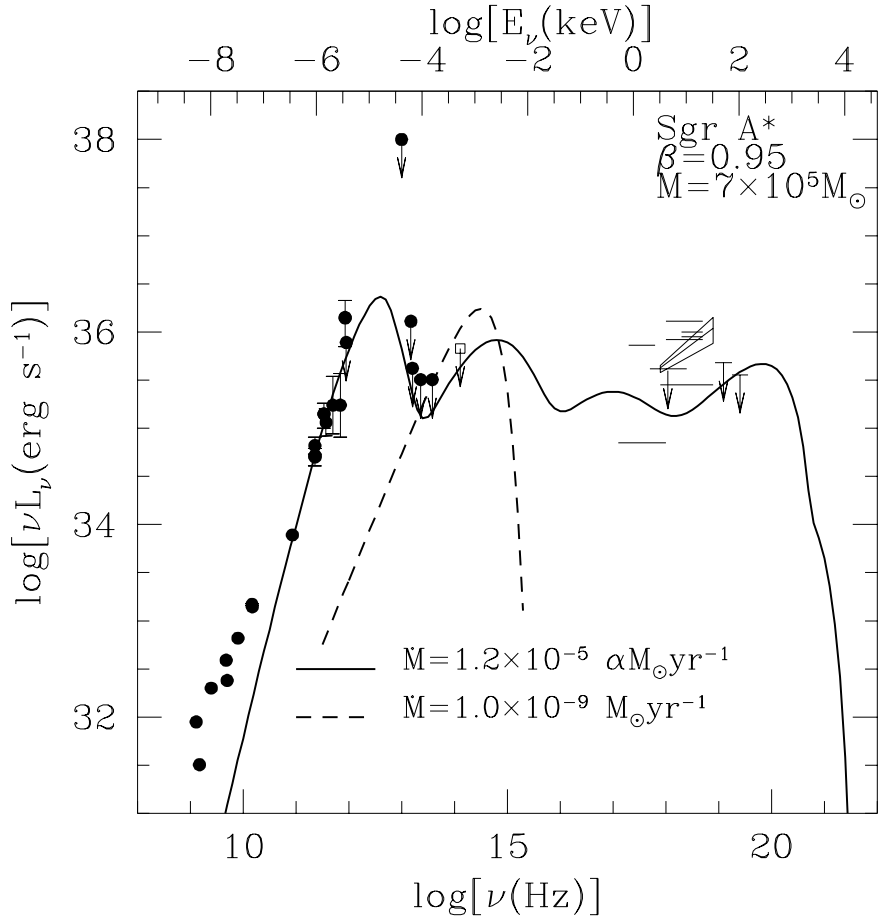

Figure 1: Fig. 1. The solid line shows the spectrum of Sgr A* corresponding to an advection-dominated accretion flow with $M = 7 \times 10^5 M_\odot$ and $\dot{M} = 1.2 \times 10^{-5} \alpha M_\odot \, \mathrm{yr}^{-1}$ (Narayan et al. 1995a). The spectrum fits the observed measurements and upper limits quite well. The dashed line corresponds to a thin accretion disk model for the same $M$ and with $\dot{M} = 10^{-9} M_\odot \, \mathrm{yr}^{-1}$. The spectrum is nearly blackbody in this case and does not explain either the radio/mm or X-ray observations.

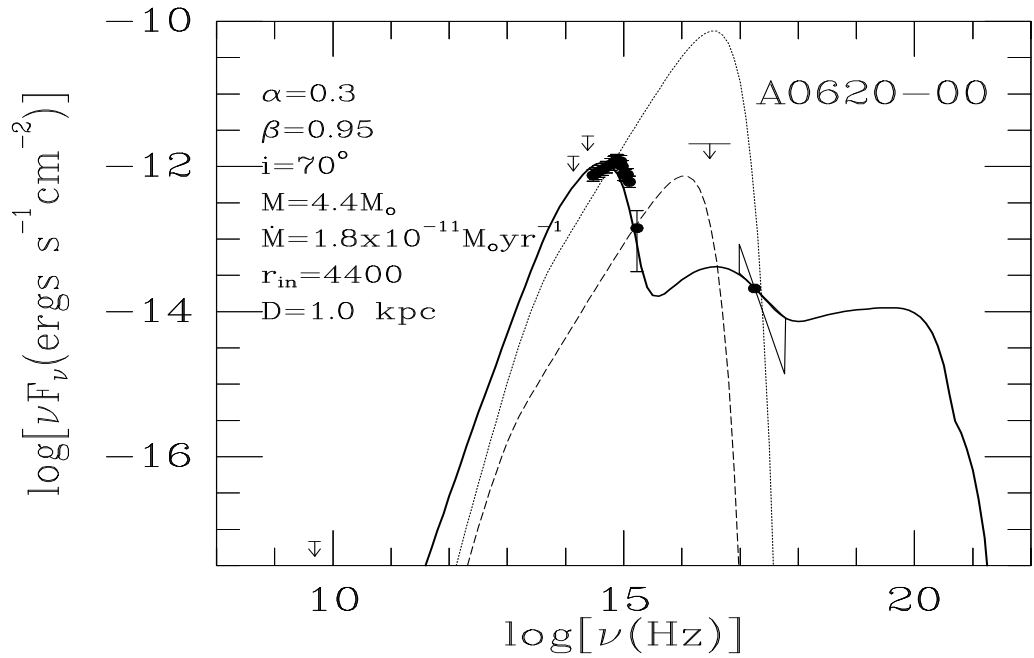

Figure 2: Fig. 2. The thick solid line is a model spectrum of the soft X-ray transient source A0620-00 in quiescence (Narayan et al. 1995b). The model assumes a thin disk down to $r = 4400$ and an advection-dominated flow from $r = 4400$ to $r = 1$ with $\alpha = 0.3$. The black hole mass is $M = 4.4 M_\odot$ and the accretion rate is $\dot{M} = 1.8 \times 10^{-11} M_\odot \, \text{yr}^{-1}$. The dotted and dashed lines are two pure thin disk models with $\dot{M} = 3 \times 10^{-12}$, $3 \times 10^{-14}$ $M_\odot \, \text{yr}^{-1}$ respectively. These models are unable to fit the data.

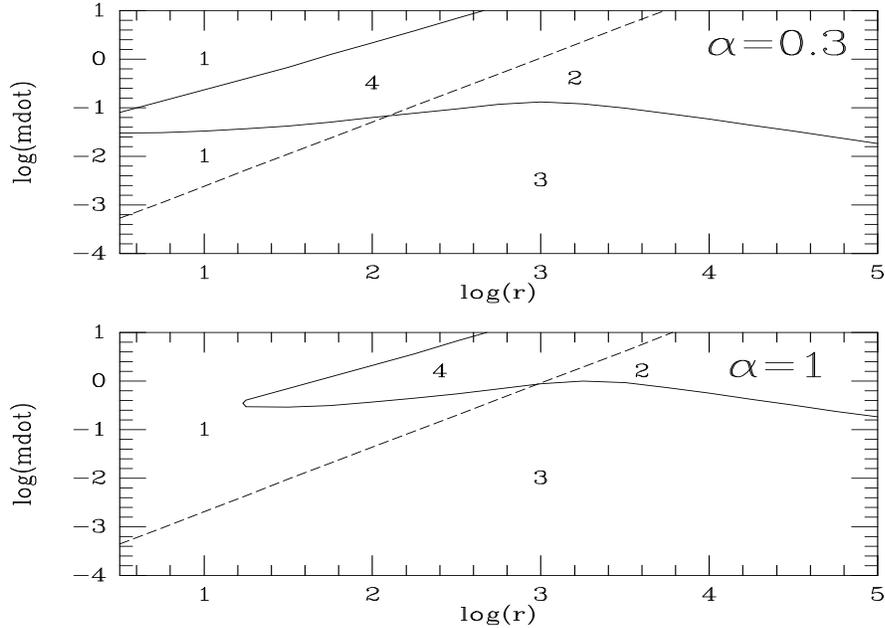

Figure 3: Fig. 3. Based on Chen et al. (1995). The upper panel shows, for $\alpha = 0.3$, the regions in the $r\dot{m}$ plane where the various known equilibrium flows are allowed. The calculations were done using the two-temperature model of Narayan & Yi (1995b) which includes cooling due to synchrotron, bremsstrahlung and Comptonization. (Note: The corresponding Fig. 2 in Chen et al. was calculated with a simpler one-temperature model which included only bremsstrahlung cooling.) The different regions are numbered $1-4$ as in Chen et al. (1995). The thin disk solution is allowed below the dashed line (above the line the accreting gas is dominated by radiation pressure and suffers from the Lightman-Eardley (1974) instability), the optically thick advection-dominated solution is allowed above the upper solid line, and the hot optically-thin advection-dominated solution, which is the focus of this article, is allowed below the lower solid line. The lower panel corresponds to $\alpha = 1$. Note that the hot optically-thin advection-dominated solution exists only for $\dot{m} \lesssim 0.3\alpha^2$. In both panels, region 3 allows two distinct equilibrium solutions, while region 4 allows no solution at all.